\documentclass[conference]{IEEEtran}
\usepackage{float}
\usepackage{cite}
\usepackage{amsmath,amssymb,amsfonts}
\usepackage{graphicx}
\usepackage{textcomp}
\usepackage{xcolor}
\usepackage{amsmath}
\usepackage{algorithm}
\usepackage{algorithmicx}
\usepackage{algpseudocode}
\usepackage{amssymb}
\usepackage{array}
\usepackage{booktabs}
\usepackage{multicol}
\usepackage{mathtools}
\usepackage{multirow}
\usepackage{url}
\usepackage{adjustbox}
\usepackage[font={small}]{caption}
\usepackage{caption}
\usepackage{subcaption}
\usepackage{colortbl}
\usepackage[
    letterpaper,
    height=9.25in,
    top=0.75in,
    bottom=1.05in,
    left=0.64in,
    right=0.64in,
    twoside=false,
    marginparwidth=0pt
]{geometry}
\begin{document}
\title{Toward Real-Time Mirrors Intelligence: System-Level Latency and Computation Evaluation in Internet of Mirrors (IoM)}

\author{
\IEEEauthorblockN{Haneen Fatima, Muhammad Ali Imran, Ahmad Taha, Lina Mohjazi}
\IEEEauthorblockA{James Watt School of Engineering, University of Glasgow, Glasgow, G12 8QQ, UK\\
Email: \{haneen.fatima, muhammad.imran,  ahmad.taha, lina.mohjazi\}@glasgow.ac.uk}
}

\maketitle

\begin{abstract}
The Internet of Mirrors (IoM) is an emerging IoT ecosystem of interconnected smart mirrors designed to deliver personalised services across a three-tier node hierarchy spanning consumer, professional, and hub nodes. Determining where computation should reside within this hierarchy is a critical design challenge, as placement decisions directly affect end-to-end latency, resource utilisation, and user experience. This paper presents the first physical IoM testbed study, evaluating four computational placement strategies across the IoM tier hierarchy under real Wi-Fi and 5G network conditions. Results show that offloading classification to higher-tier nodes substantially reduces latency and consumer resource load, but introduces network overhead that scales with payload size and hop count. No single strategy is universally optimal: the best choice depends on available network, node proximity, and concurrent user load. These findings empirically characterise the computation-communication trade-off space of the IoM and motivate the need for intelligent, adaptive task placement responsive to application requirements and live ecosystem conditions.
\end{abstract}

\section{Introduction}

\begin{figure*}[!t]
    \centering
    \includegraphics[width=0.85\textwidth]{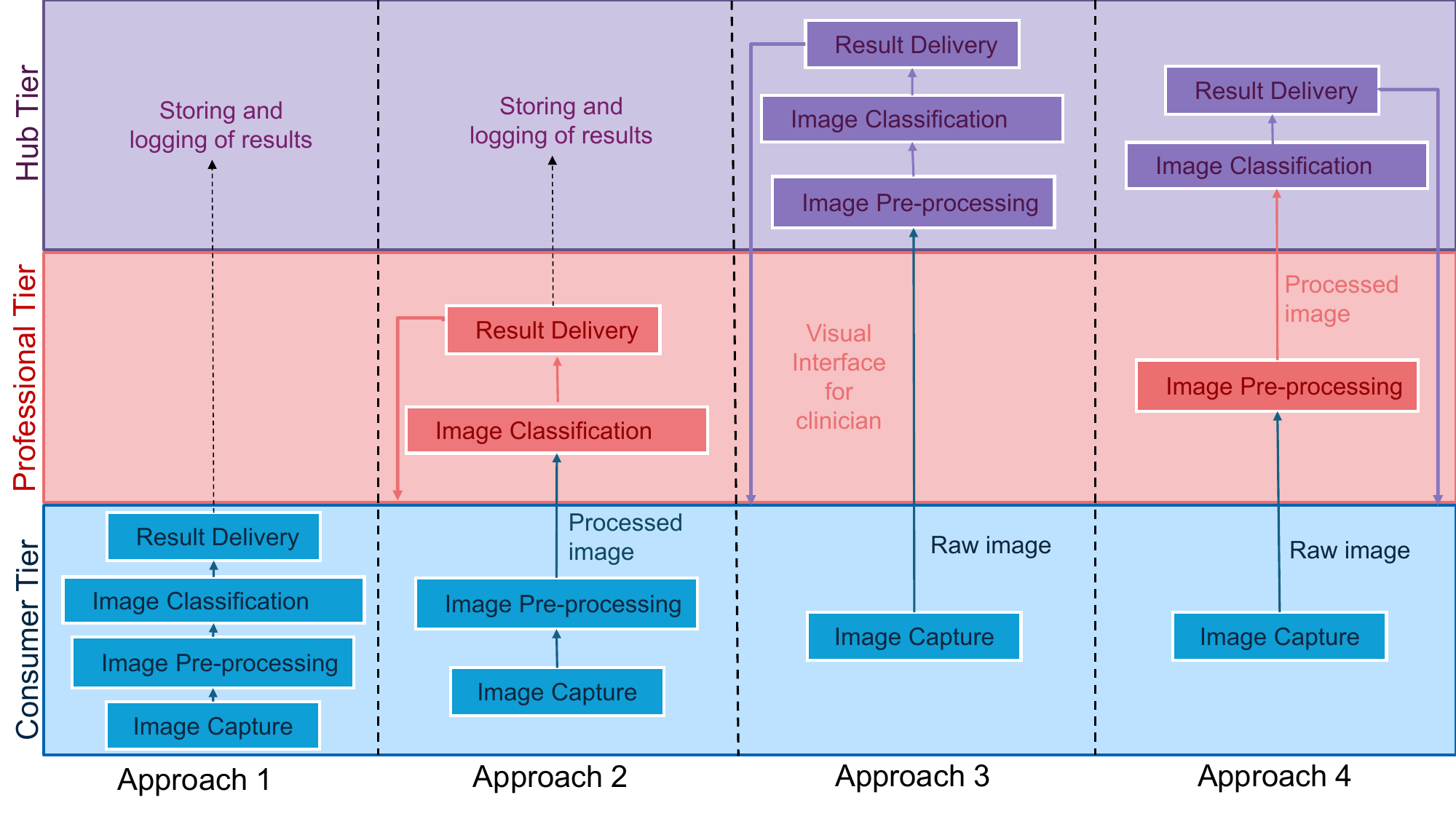} \vspace{-0.3cm}
    \caption{Computational placement strategies across the IoM tier hierarchy.}
    \label{fig:approaches} 
    \vspace{-0.5cm}
\end{figure*}

The rise of IoT has driven a shift from centralised cloud architectures toward distributed computing paradigms that position processing closer to the data source~\cite{shi_edge_2016, satyanarayanan_emergence_2017}, giving rise to the edge-fog-cloud continuum within which task placement decisions carry direct implications for latency, resource utilisation, and scalability~\cite{shahid_iot_2026, fernando_edge-fog-cloud_2025}. Within this context, the Internet of Mirrors (IoM) has been introduced as an emerging ecosystem under the umbrella of IoT, comprising of an interconnected network of smart mirrors equipped with sensing, processing, and communication 
capabilities, designed to deliver personalised health, beauty, and assisted living services through an immersive visual interface~\cite{fatima_internet_2023}. The IoM is organised as a three-tier hierarchy spanning consumer, professional, and hub nodes, each reflecting a distinct stakeholder context and level of computational resource availability. This stakeholder-driven structure gives rise to heterogeneous application requirements across tiers, making the placement of computation a critical system design consideration.

Prior work on tiered IoT architectures has established that neither fully local nor fully remote execution is universally optimal, with hierarchical placement strategies shown to reduce end-to-end latency relative to single-tier approaches~\cite{guo_energy-efficient_2019, fernando_edge-fog-cloud_2025}. This has been particularly relevant in healthcare IoT, where fog-proximate processing addresses the latency sensitivity of diagnostic applications~\cite{ahmadi_fog-based_2021}. The choice of network technology further complicates placement decisions, empirical comparisons have shown that 5G sustains lower 
latency and fewer outliers under concurrent load compared to Wi-Fi~\cite{arendt_distributed_2024}, while latency sensitivity has been demonstrated on live 5G edge testbeds~\cite{cao_evaluating_2021} and in health-adjacent interactive systems~\cite{abdellatif_edge_2019}. Multi-radio frameworks 
have also identified that task allocation across available network technologies directly affects service latency~\cite{ali_multiradio_2024}. However, these works are grounded in simulation and analytical modelling rather than physical deployment, and do not address the stakeholder-driven tier structure of the IoM, where placement decisions carry direct implications for user experience across scenarios ranging from real-time cosmetic augmentation at the consumer tier to AI-driven clinical diagnostics at the hub~\cite{fatima_enabling_2025}.

An initial feasibility study comparing standalone edge and cloud deployments for a single IoM node identified network latency as a critical bottleneck and suggested that a hybrid placement approach would be necessary to optimise the computation-communication trade-off in larger, interconnected systems~\cite{fatima_latency_2024}. That study was limited to a simulated, single-node configuration and did not examine how different placement strategies perform across a real IoM hierarchy under real network conditions. To the best of our knowledge, no study has placed the IoM on the computing continuum or implemented its heterogeneous tiers on a physical testbed. This work addresses this gap through the first physical IoM testbed study, employing our existing dental smile analysis use case~\cite{baidachna_mirror_2024} to systematically evaluate four computational placement strategies across the IoM tier hierarchy under real Wi-Fi and 5G conditions. Due to the diversity of IoM applications, understanding where IoM workloads sit on the computing continuum, and how placement decisions interact with real network conditions, is a necessary foundation for IoM system design.

The main contributions of this work are as follows:
\begin{itemize}
    \item The first deployment and evaluation of a physical IoM testbed
    \item The first systematic comparison of computational placement strategies across the 
    IoM architecture
    \item An empirical analysis of the computation-communication trade-off in IoM systems 
    under both Wi-Fi and 5G network conditions
\end{itemize}

\section{System Design}
\subsection{IoM Architecture}

The IoM is organised as a three-tier node hierarchy, with each tier reflecting distinct computational capabilities and deployment contexts~\cite{fatima_internet_2023}. 
\begin{itemize}
    \item \textbf{Hub nodes} represent the highest tier, comparable in capability to enterprise edge servers or institutional data infrastructure. 
    \item \textbf{Professional nodes} occupy the intermediate tier, offering medium-capability processing suited to professional service environments such as dermatology clinics, beauty salons, and pharmacy settings. 
    \item \textbf{Consumer nodes} form the lowest tier, providing basic computational capability for widespread personal or small-retail deployment. 
\end{itemize}
This stakeholder-driven structure means that placement of computation is not merely a technical optimisation problem but carries direct implications for service quality, data locality, and user experience across diverse deployment contexts.


\subsection{Application Workload: Dental Smile Analysis Pipeline}

Dental smile analysis~\cite{baidachna_mirror_2024} is used as the representative IoM application workload throughout this study. The pipeline comprises four sequential stages that are consistent across all placement strategies, differing only in which tier executes each stage:

\begin{itemize}
    \item \textbf{Capture:} A 600$\times$600 pixel JPEG image is acquired using a face detection algorithm, via a webcam attached to the Consumer node.
    \item \textbf{Pre-processing:} The captured image undergoes segmentation and normalisation to produce a feature representation suitable for inference.
    \item \textbf{Classification:} A MobileNetV2-based CNN classifies the smile condition on the pre-processed input. This is the most computationally intensive stage of the pipeline.
    \item \textbf{Result Delivery:} A structured JSON result is returned to the Consumer node for display on the smart mirror interface.
\end{itemize}

This pipeline is well-suited as an initial testbed workload because it spans a range of computational intensities, from lightweight capture and pre-processing to CNN inference. Alongside this, it involves data payloads of varying sizes at each inter-tier transfer point, making the computation-communication trade-off observable across all four placement strategies.

\subsection{Computational Placement Strategies}
\label{sec:strategies}
\begin{table*}[!t]
\caption{Summary of Computational Placement Strategies}
\label{tab:strategies}
\centering
\renewcommand{\arraystretch}{1.4}
\begin{tabular}{|p{2.1cm}|p{4.2cm}|p{2.7cm}|p{1.6cm}|p{4.3cm}|}
\hline
\textbf{Strategy} & \textbf{Stage-to-Tier Mapping} & \textbf{Payload Transferred} & \textbf{Network Hops} & \textbf{Deployment Scenario} \\
\hline
Consumer-only & Stages 1-4: Consumer \newline log upload: Consumer $\rightarrow$ Hub & JSON log ($\sim$few KB) & 1 & Patient uses the mirror at home, inference is done locally. Only logs or summary results are uploaded for the dentist to review later. \\
\hline
Professional-offload & Stage 1-2: Consumer \newline Stage 3: Professional \newline Stage 4: Professional $\rightarrow$ Consumer \newline log: Consumer $\rightarrow$ Hub & Pre-processed image ($\sim$200-500\,KB)  & 3 & Clinic mirror: A walk-up mirror in a pharmacy or clinic waiting area sends pre-processed images to a nearby dental workstation for quicker chair-side assessment. \\
\hline
Hub-offload & Stage 1: Consumer $\rightarrow$ Hub \newline Stages 2-3: Hub \newline Stage 4: Hub $\rightarrow$ Consumer & Raw image ($\sim$2-3\,MB) & 2 & Large clinic setup: Several mirrors send images to a central practice server suitable when many patients are checked in different rooms/nearby clinics.  \\
\hline
Tiered-distributed & Stage 1-2: Consumer \newline Stage 3: Professional $\rightarrow$ Hub \newline Stage 4: Hub $\rightarrow$ Consumer & Raw image then segmented image ($\sim$2-3\,MB then $\sim$500\,KB) & 3 & Mirror at the chair captures the image, professional node performs detailed analysis, and the hub stores patient records for longitudinal tracking.\\
\hline
\end{tabular}
\end{table*}

Four placement strategies are defined, each representing a distinct point on the computing continuum and mapping to a realistic IoM deployment scenario, as illustrated in Fig.~\ref{fig:approaches}.
These four strategies span the range from fully local to fully distributed execution, and together they characterise the computation-communication trade-off space of the IoM tier hierarchy. A summary of the stage-to-tier mapping, network hops, and transmitted payloads for each strategy is provided in Table~\ref{tab:strategies}.

\section{Experimental Setup}

\subsection{Testbed Hardware} 
The physical testbed implements the three-tier IoM hierarchy using representative embedded and edge computing hardware comprised of five active computing nodes. 

The consumer tier comprises three Raspberry Pi 4 Model B devices (4\,GB RAM, quad-core Cortex-A72 CPU), each connected to a USB webcam, representing resource-constrained consumer nodes. 

The professional and hub tiers are each implemented on an NVIDIA Jetson Xavier NX (8\,GB RAM, 6-core Carmel ARM CPU, 384-core Volta GPU with 512 CUDA cores), reflecting the higher processing capacity available in clinical or centralised deployment settings. CNN inference at the professional and hub tiers executes on the Volta GPU, while all consumer-tier processing runs on general-purpose CPU cores. 

\begin{figure}[!b]
    \centering
    \includegraphics[width=0.5\textwidth]{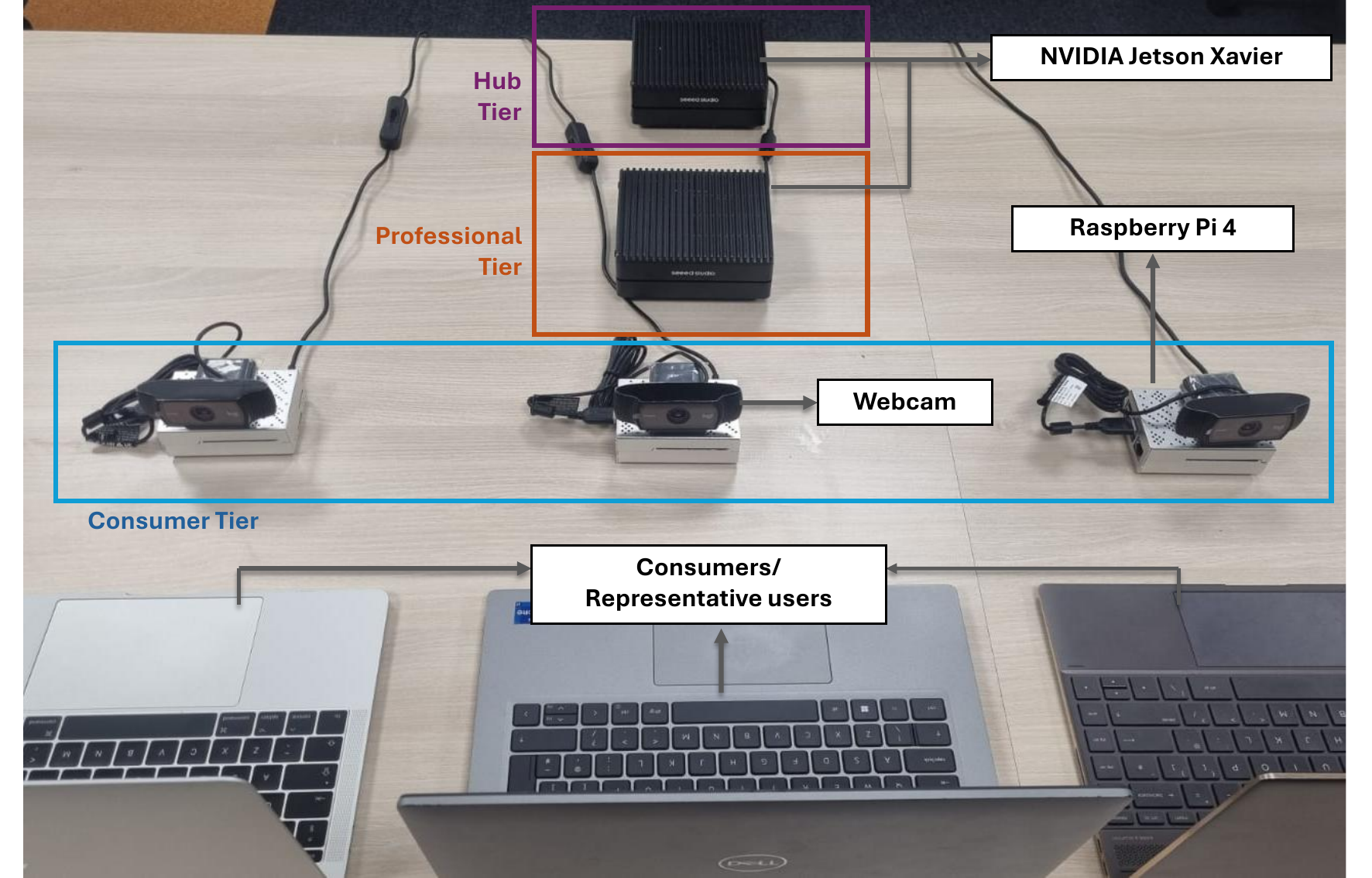}
    \caption{Physical IoM testbed configuration showing consumer, professional, and hub nodes}
    \label{fig:testbed} 
\end{figure}

Network connectivity is provided via a local Wi-Fi router and a public 5G CPE (Customer Premises Equipment). The physical testbed setup is shown in Figure~\ref{fig:testbed}.

\subsection{Workload and Experimental Design}
The dental smile analysis pipeline described in Section~\ref{sec:strategies} is used as the application workload throughout. Input images are 600$\times$600 pixel JPEG photographs. Pre-processing applies image segmentation and normalisation, and classification is performed using a MobileNetV2-based CNN. To control for image variability as an experimental factor, a fixed set of captured images is reused across trials, isolating computational and network behaviour as the independent variables.



Experiments are conducted across a full factorial design covering all combinations of the following variables:
\begin{itemize}
    \item \textbf{Placement strategy:} four approaches (Consumer-only, Professional-offload, Hub-offload, Tiered-distributed)
    \item \textbf{Network condition:} Wi-Fi and 5G
    \item \textbf{Concurrent consumers:} 1, 2, and 3 simultaneous consumer nodes
\end{itemize}
Each combination is repeated for 25 independent trials, yielding a total of 600 trials (4 strategies $\times$ 2 networks $\times$ 3 concurrency levels $\times$ 25 repetitions).

\subsection{Metrics}
Prior to each experiment, device clocks are synchronised using \texttt{timedatectl} with NTP enabled on all nodes to ensure consistent cross-device timestamping for accurate metric collection. Performance is characterised through four key metrics,:
\begin{itemize}
    \item \textbf{End-to-end (E2E) latency}: total elapsed time from image capture to result display at the consumer node, decomposed by pipeline stage.
    \item \textbf{Per-stage latency}: individual processing time for each pipeline stage.
    \item \textbf{Network transfer latency}: per-hop communication time between tiers, capturing the cost of each inter-tier data transfer.
    \item \textbf{Resource utilisation}: combined CPU and GPU utilisation at each active tier node, collected using the \texttt{psutil} library, GPU utilisation is recorded on Jetson devices where applicable.
    \item \textbf{Scalability}: degradation in E2E and network latency as concurrent consumer count increases from 1 to 3, including variance metrics.
\end{itemize}

\section{Results and Discussion}
The results presented in this section are structured to progressively characterise the computation-communication trade-off across the four placement strategies, beginning with overall end-to-end latency, decomposing network contributions by hop, examining resource distribution across tiers, and assessing scalability under concurrent load. Together, these perspectives provide the empirical grounding to evaluate where IoM applications sit on the computing continuum and what this means for deployment.

\subsection{End-to-End Latency Breakdown}
\label{sec:e2e}

\begin{figure}[h]
    \centering \vspace{-0.3cm}
    \includegraphics[width=1\linewidth]{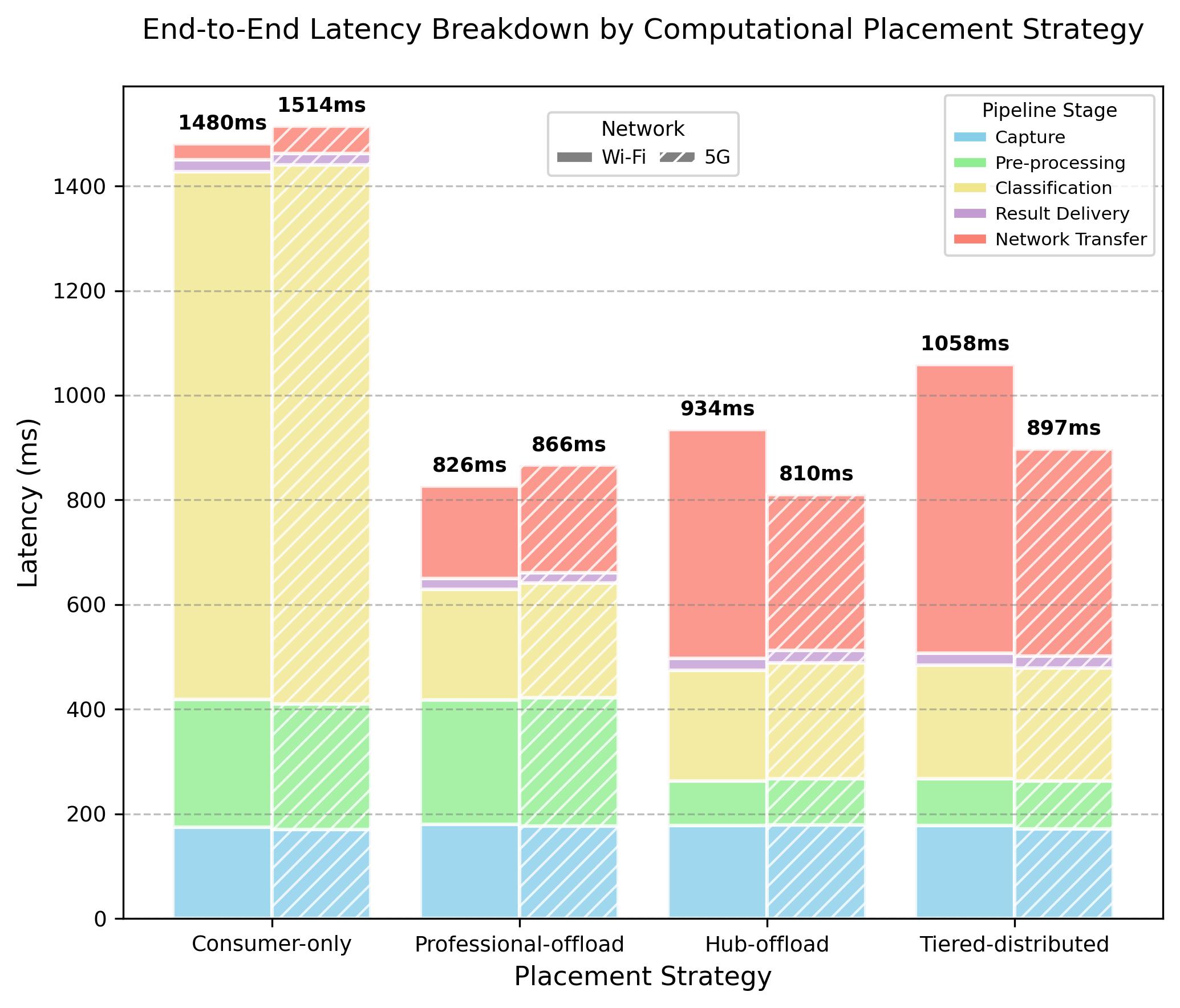}
    \caption{End-to-end latency decomposed by pipeline stage and network technology across the four computational placement strategies.}
    \label{fig:e2elatency} 
    \vspace{-0.25cm}
\end{figure}

Figure~\ref{fig:e2elatency} presents E2E latency decomposed by pipeline stage and network technology. Consumer-only exhibits the highest latency on both Wi-Fi (1,481~ms) and 5G (1,515~ms), with Classification alone accounting for 68\% of total E2E time (1,007~ms). This reflects the fundamental constraint of executing CNN inference sequentially on general-purpose CPU cores at the Consumer tier meaning that the processing burden is not a function of network conditions but of local hardware capacity, which is why Wi-Fi and 5G results are nearly identical for this approach.

Offloading classification to higher tiers substantially reduces this bottleneck. Professional-offload achieves the lowest Wi-Fi latency (826~ms), reducing Classification time to around a quarter of the end-to-end latency by executing inference on a higher tier. This comes at the cost of 177~ms of network transfer overhead (21\% of E2E). On 5G, Hub-offload becomes the fastest approach (810~ms), benefiting from reduced raw image transfer latency compared to Wi-Fi (298~ms vs 438~ms). Tiered-distributed incurs the highest network overhead of all offloading approaches (52\% of E2E on Wi-Fi, 44\% on 5G), reflecting its three-hop routing structure.


These results establish that offloading relieves the classification bottleneck but substitutes it with a network cost that grows with payload size and hop count. The degree to which this substitution is beneficial depends on both the network technology available and the tier to which processing is offloaded. Understanding exactly where this network overhead originates requires examining each transfer hop individually.

\subsection{Network Transfer Analysis}

\begin{figure}[h]
    \centering \vspace{-0.5cm}
    \includegraphics[width=1\linewidth]{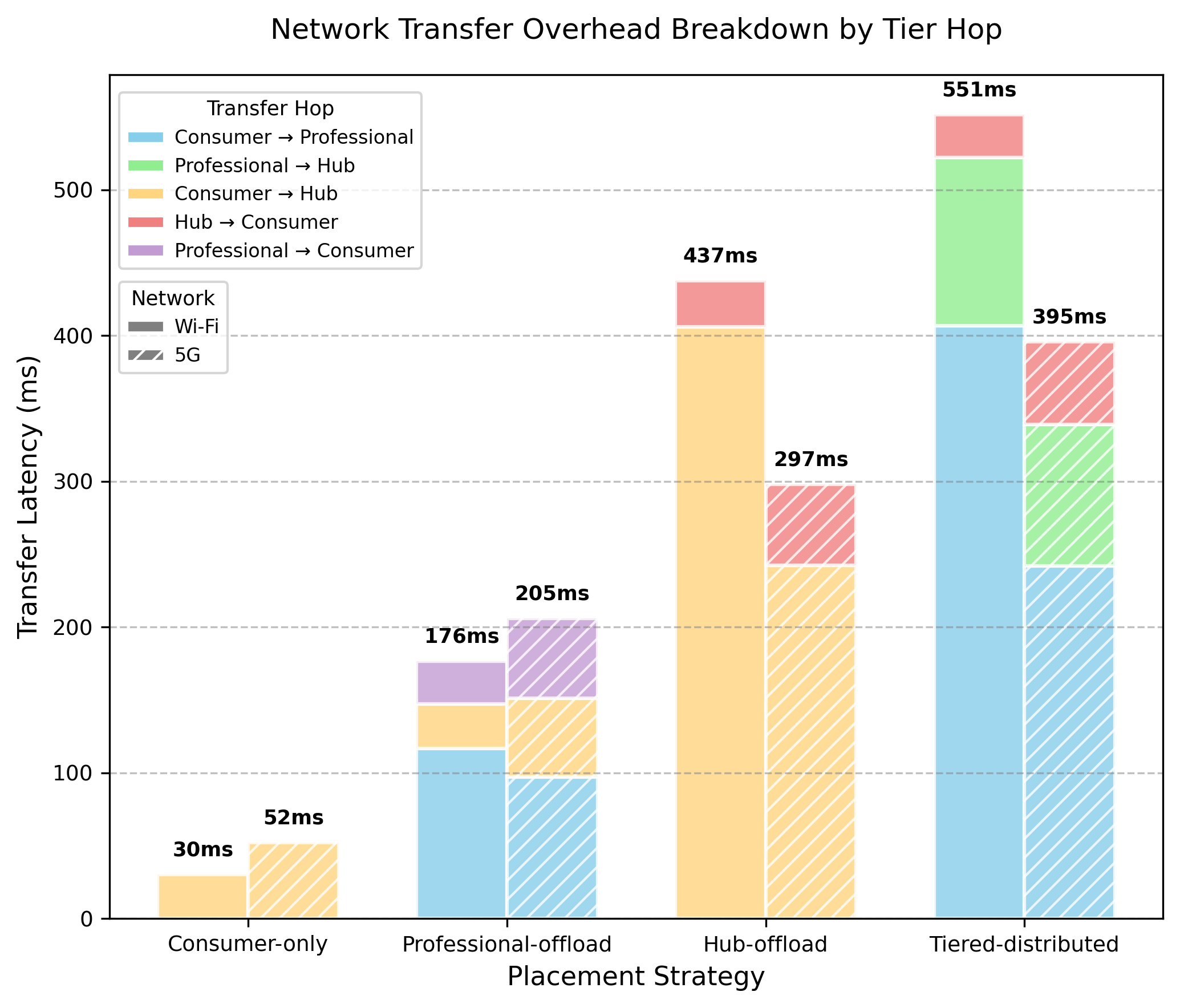}
    \caption{Network transfer overhead decomposed by inter-tier hop direction for each placement strategy.}
    \label{fig:tier_hop} 
    \vspace{-0.25cm}
\end{figure}

Figure~\ref{fig:tier_hop} decomposes total network transfer latency by hop direction, revealing the mechanisms behind the Wi-Fi versus 5G patterns observed in Figure~\ref{fig:e2elatency}. Consumer-only incurs minimal transfer cost (30~ms Wi-Fi, 52~ms 5G), uploading only a small log file. The slight 5G penalty here is notable: for a payload of just a few KB, the cellular access overhead exceeds any bandwidth benefit, making 5G marginally worse for this approach.

Professional-offload shows moderate total overhead (177~ms Wi-Fi, 206~ms 5G), with the Consumer$\rightarrow$Professional hop carrying a pre-processed feature vector of 200--500~KB (117~ms Wi-Fi, 97~ms 5G). The relatively small 5G benefit at this hop reflects the moderate payload size, 5G's bandwidth advantage is present but limited. Hub-offload and Tiered-distributed incur substantially higher costs due to raw JPEG payloads of 2--3~MB at the first hop. The Consumer$\rightarrow$Hub hop in Hub-offload costs 406~ms on Wi-Fi but only 242~ms on 5G, a 40\% reduction which directly explains Hub-offload's 124~ms E2E improvement on 5G. Tiered-distributed's Consumer$\rightarrow$Professional hop carries the same raw image (407~ms Wi-Fi, 242~ms 5G), with a further Professional$\rightarrow$Hub hop adding 116~ms (Wi-Fi) or 97~ms (5G), yielding a total network latency of 551~ms on Wi-Fi versus 396~ms on 5G.

The Wi-Fi-to-5G network latency difference therefore scales with payload size and hop count. Consumer-only sees a small penalty (+22~ms), Professional-offload a marginal increase (+29~ms), while Hub-offload and Tiered-distributed realise reductions of $-$140~ms and $-$155~ms respectively. For IoM deployments, this finding has a direct implication, 5G's bandwidth advantage only becomes meaningful when there is sufficient data volume to exploit it. For small payloads, the cellular access overhead dominates, whereas for large raw image transfers across multiple hops, 5G delivers a compounding benefit at each hop. For the IoM, this establishes that the value of 5G connectivity is not uniform across deployment scenarios but is specifically significant for hub-centric or fully distributed strategies, and largely irrelevant for consumer-local or near-edge approaches. 

\subsection{Resource Utilisation}
\label{subsec:resource}

\begin{figure}[!ht]
    \centering  \vspace{-0.3cm}
    \includegraphics[width=1\linewidth]{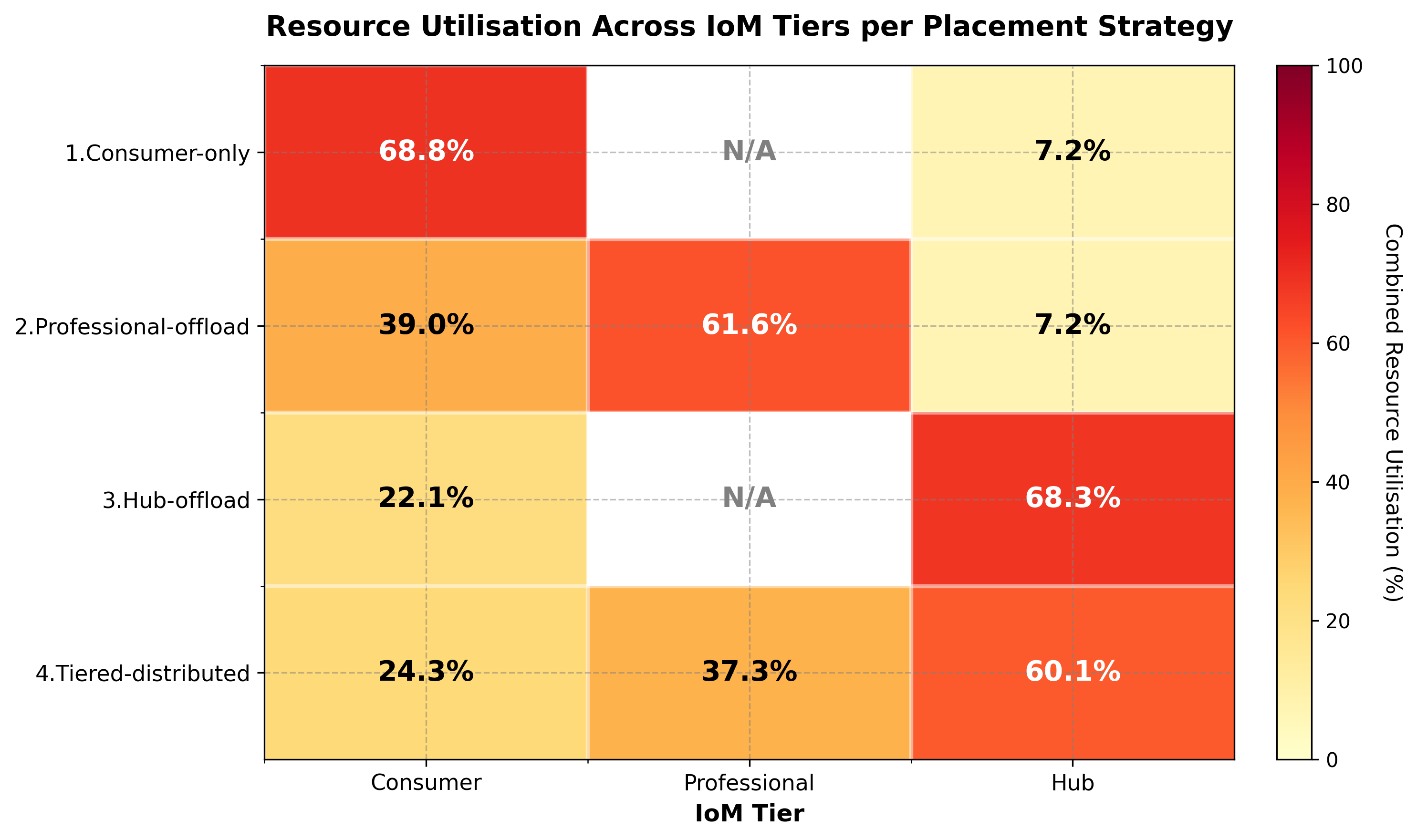}
    \caption{Combined Resource utilisation (\%) across IoM tiers per placement strategy.}
    \label{fig:resource_util}
    \vspace{-0.2cm}
\end{figure}

Fig.~\ref{fig:resource_util} presents resource utilisation across IoM tiers as a heatmap, showing how each placement strategy distributes the computational burden of the use case pipeline. Consumer-only concentrates 68.8\% of utilisation at the consumer tier, with the professional tier unused and the hub at 7.2\% for log storage only. The offloading strategies redistribute this clearly, Professional-offload reduces consumer utilisation to 39.0\% while shifting 61.6\% to the professional tier. Whereas Hub-offload concentrates 68.3\% at the hub with consumer utilisation falling to 22.1\% and the professional tier again unused; and Tiered-distributed spreads load across all three tiers (consumer 24.3\%, professional 37.3\%, hub 60.1\%).

These patterns carry implications for the IoM that extend beyond latency. The 68.8\% consumer utilisation in Consumer-only places the smart mirror device under sustained compute load, with direct consequences for the device's ability to handle concurrent tasks which are envisioned to be numerous on a real IoM. Reducing this to 22-39\% through offloading substantially improves the operational headroom of the consumer node. From an architectural standpoint, the heatmap also validates the IoM's tier design, each tier absorbs the workload it was designed for, with lightweight tasks remaining at the consumer edge and intensive inference concentrated at higher-capacity nodes. This tier-appropriate load distribution is not only a computational latency optimisation it directs the way for sustainable, multi-application IoM deployments where the consumer mirror must remain responsive across a range of simultaneous tasks and applications.

\subsection{Scalability Under Concurrent Load}
\label{subsec:scalability}

\begin{figure}[h]
    \centering \vspace{-0.3cm}
    \includegraphics[width=1\linewidth]{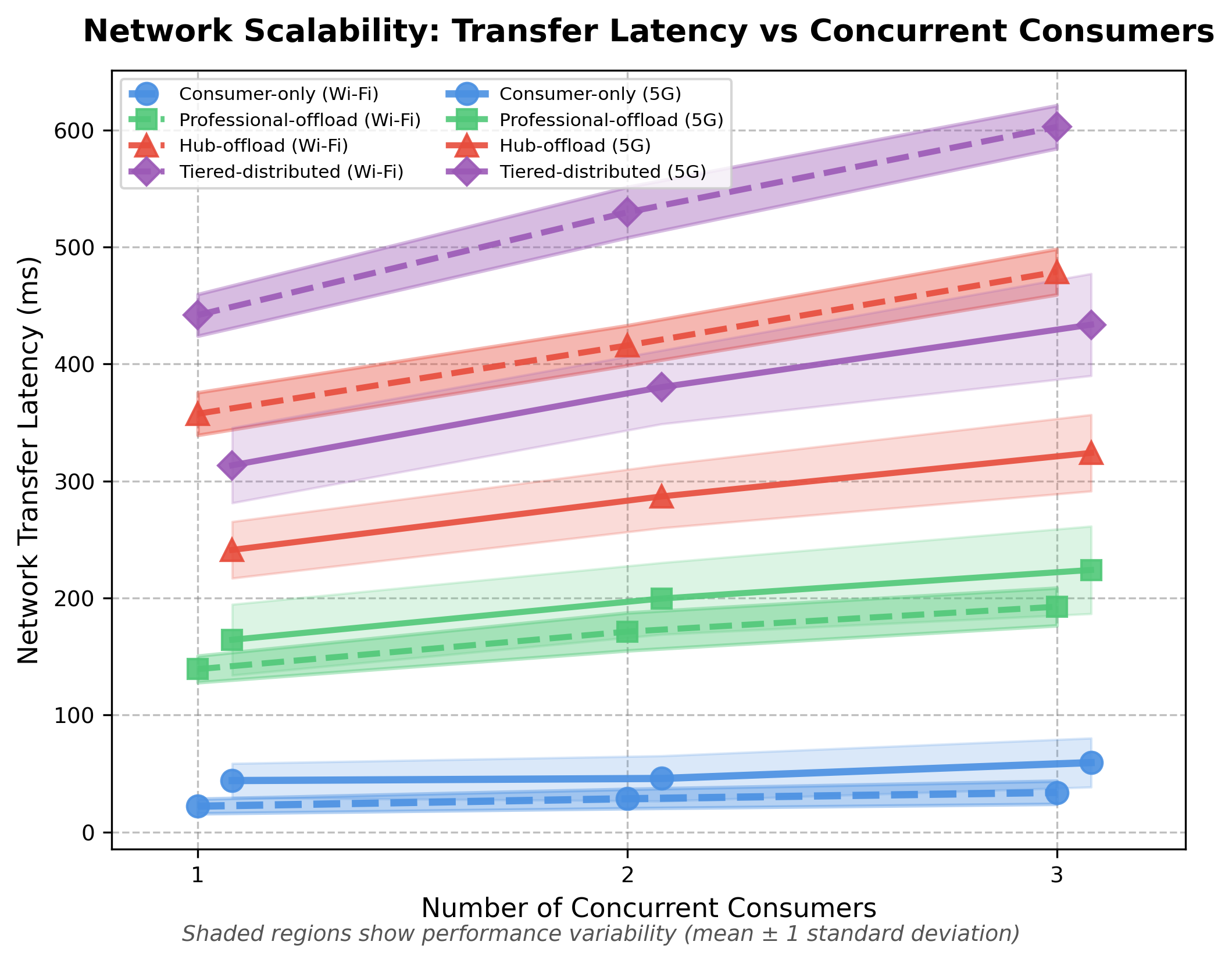}
    \caption{Network transfer latency scalability analysis under concurrent consumer load.}
    \label{fig:scalability}
    \vspace{-0.3cm}
\end{figure}

Fig.~\ref{fig:scalability} presents network transfer latency under increasing concurrent consumer load. Preliminary analysis of compute latency confirmed nearly identical degradation slopes and variance bands across Wi-Fi and 5G for all strategies, establishing that compute behaviour is network-independent, end-to-end differences between network types are therefore attributable to transfer overhead alone.

Consumer-only remains flat and minimal throughout (22\,ms$\rightarrow$34\,ms on Wi-Fi), consistent with its negligible transfer activity. Its compute degradation is the steepest of all strategies, rising from 1,198\,ms to 1,578\,ms at three concurrent consumers (190\,ms per additional consumer), with variance widening substantially reflecting queuing and resource contention as the consumer CPU saturates under concurrent inference requests. Professional-offload, by contrast, degrades at only 82\,ms per consumer with coefficient of variation (CV) shrinking, as GPU-capable higher tiers handle concurrent requests with substantially less contention.

Hub-offload and Tiered-distributed exhibit steep network gradients on Wi-Fi as concurrent requests amplify inter-tier congestion, on 5G both show slighty shallower gradients, demonstrating that 5G limits congestion growth under load. However, the shaded variance bands are visibly wider on 5G for these strategies, 5G reduces mean latency growth but introduces greater trial-to-trial variability, likely reflecting the less deterministic nature of cellular conditions compared to a controlled local Wi-Fi environment.

\subsection{Deployment Implications}
\label{subsec:implications}

Across the four subsections, a consistent picture emerges of how the computation-communication trade-off manifests across different IoM deployment contexts.
Table~\ref{tab:strategies} maps each strategy to a scenario, and the results validate and refine those mappings empirically.
\begin{itemize}
    \item \textbf{Consumer-only} keeps all processing local, minimising network dependency and making it suitable for private, offline deployments such as a home smart mirror. However, its high consumer tier utilisation and steep latency degradation under concurrent load make it unsuitable for any multi-user scenario. Its growing variance under load is particularly problematic for the IoM, where user experience consistency is as important as mean latency.
    \item \textbf{Professional-offload} emerges as a balanced strategy. It achieves the lowest Wi-Fi latency, a stable variance under load, and relieves the consumer node sufficiently for responsive concurrent operation. It is the recommended strategy for clinic or pharmacy settings where a professional node is co-located and Wi-Fi is the available network.
    \item \textbf{Hub-offload} is the strongest performer on 5G, where its raw image transfer cost is substantially reduced. It is well-suited to large multi-room clinic deployments with reliable high-bandwidth connectivity, where a centralised server can handle simultaneous requests from multiple mirrors without requiring a professional node at every point of care.
    \item \textbf{Tiered-distributed} makes full use of the IoM hierarchy and is the most appropriate strategy where longitudinal record-keeping is a priority alongside real-time analysis. Its network overhead is the highest of all strategies, but its load distribution across all three tiers means no single node is disproportionately burdened, making it suitable for large scale IoM deployments where all tiers are consistently available and 5G connectivity is present.
\end{itemize}

The results reveal a consistent computation-communication trade-off that manifests differently across IoM deployment contexts. Consumer-only minimises network dependency but is constrained by consumer hardware capacity and degrades unpredictably under load, Professional-offload and Hub-offload reduce the compute burden on the consumer node but introduce network overhead that grows with payload size and concurrency, and Tiered-distributed distributes load most evenly across the hierarchy but at the cost of the highest cumulative transfer overhead. The optimal choice shifts depending on the network available, the proximity and availability of higher-tier nodes, the computational intensity of the application, and the number of concurrent users, all of which vary across real IoM deployments. This context-dependence is itself a key finding of this work, it establishes that the IoM ecosystem cannot be adequately served by a fixed placement policy, and that intelligent, dynamic optimisation of task placement, responsive to application type, task characteristics, and live ecosystem conditions is a necessary direction for future IoM system design.
\vspace{-0.2cm}

\section{Conclusion}

This paper presented the first physical IoM testbed study, evaluating four computational placement strategies across a three-tier IoM hierarchy under real Wi-Fi and 5G network conditions, using dental smile analysis as a representative application workload. The results demonstrate that offloading intensive tasks to higher-tier nodes substantially reduces end-to-end latency and consumer resource utilisation, but introduces network overhead that grows with payload size, hop count, and concurrent load, with no single strategy optimal across all conditions. 
The network analysis further reveals that 5G's benefit is payload-dependent, providing meaningful gains only for strategies involving large inter-tier transfers, while introducing greater trial-to-trial variability compared to Wi-Fi.
These findings establish that the IoM ecosystem cannot be adequately served by a fixed placement policy, and motivate the need for intelligent, adaptive task placement that responds to application requirements, task characteristics, and live network and resource conditions across the hierarchy. Addressing this challenge is an important direction for future IoM system design.

\vspace{-0.4cm}

\bibliographystyle{ieeetr}
\bibliography{ref}

\end{document}